\renewcommand{\vec}[1]{ {\mathbf #1} }

\newcommand{\eps}{ \epsilon^{ijk}}
\newcommand{\gth}{ \Theta}

\newcommand{\Vb}{ \vec b}
\newcommand{\Vr}{ \vec r}
\newcommand{\Vbp}{ \vec b'}
\newcommand{\Vrp}{ \vec r'}
\newcommand{\Vv}{ \vec v}

\newcommand{\Veone}{ \vec e^1}
\newcommand{\Vetwo}{ \vec e^2}
\newcommand{\Vethree}{ \vec e^3}

\newcommand{\Vsigma}{\boldsymbol\sigma}

\newcommand{\VLambda}{\boldsymbol\Lambda}
\newcommand{\VLambdar}{\boldsymbol\Lambda_{rot}}
\newcommand{\VLambdab}{\boldsymbol\Lambda_{boost}}
\newcommand{\rx}{ r_x}
\newcommand{\ry}{ r_y}
\newcommand{\rz}{ r_z}
\newcommand{\rone}{ r^1}
\newcommand{\rtwo}{ r^2}
\newcommand{\rthree}{ r^3}
\newcommand{\ri}{ r^i}
\newcommand{\rxp}{ r'^1}
\newcommand{\ryp}{ r'^2}
\newcommand{\rzp}{ r'^3}

\newcommand{\bx}{ b_x}
\newcommand{\by}{ b_y}
\newcommand{\bz}{ b_z}
\newcommand{\bone}{ b^1}
\newcommand{\btwo}{ b^2}
\newcommand{\bthree}{ b^3}
\newcommand{\bi}{ b^i}

\newcommand{\bxp}{ b'^1}
\newcommand{\byp}{ b'^2}
\newcommand{\bzp}{ b'^3}

\newcommand{\tone}{ \theta^1}
\newcommand{\ttwo}{ \theta^2}
\newcommand{\tthree}{ \theta^3}
\newcommand{\ti}{ \theta^i}
\newcommand{\etone}{ \eta^1}
\newcommand{\ettwo}{ \eta^2}
\newcommand{\etthree}{ \eta^3}
\newcommand{\eti}{ \eta^i}

\newcommand{\GsruL}{{\cal G}^L_{rot}(\vec u)}

\newcommand{\GsruR}{{\cal G}^R_{rot}(\vec u)}

\newcommand{\Gr}{ G_{rot}(\Vr)}
\newcommand{\Grb}{ G_{rot}}
\newcommand{\Grp}{ G_{rot}(\Vrp)}

\newcommand{\GsbuL}{{\cal G}^L_{boost}(\vec u)}
\newcommand{\GsbuR}{{\cal G}^R_{boost}(\vec u)}

\newcommand{\Gbb}{G_{boost} }
\newcommand{\Gb}{G_{boost}(\Vb) }
\newcommand{\Gbp}{G_{boost}(\Vbp) }

\newcommand{\pTheta}{\partial_{\Theta}}

\newcommand{\pv}{ \partial_{{\rm v}}}
\newcommand{\pmcV}{ \partial_{\mcV}}
\newcommand{\mcV}{\mathcal{V}}

\documentclass[11pt,a4paper]{article}
\usepackage[latin1]{inputenc}
\usepackage{amsmath}
\usepackage{amsfonts}
\usepackage{amssymb}
\begin{document}
\small
\title{Lorentz Transformations}

\author{Bernard R. Durney}
\date{2377 Route de Carc\'{e}s, F-83510 Lorgues, France durney@physics.arizona.edu} 

\maketitle

\bigskip

\smallskip\textbf{Abstract}.
This paper describes a particularly didactic and transparent derivation of basic
properties of the Lorentz group. The generators for rotations and boosts along an arbitrary direction, as well as their commutation relations, are written as functions of the unit vectors that define the axis of rotation or the direction of the boost (an approach that can be compared with the one that in electrodynamics, works with the electric and magnetic fields instead of the Maxwell stress tensor). For finite values of the angle of rotation or the boost's velocity (collectively denoted by $\mathcal{V})$, the existence of an exponential expansion for the coordinate transformation's matrix, $\mathcal{M},$ in terms of $G \mathcal{V}$ with $G$ being the generator, requires that the matrix's derivative with respect to $\mathcal{V}$ be equal to $G\mathcal{M}$. This condition can only be satisfied if the transformation is additive as it is indeed the case for rotations, but not for velocities. If it is assumed, however, that for boosts such an expansion exists, with $\mathcal{V} =\mathcal{V}(v)$, $v$ being the boodt's velocity, and if the above condition is imposed on the boost's matrix, then its expression in terms of hyperbolic $\cosh(\mathcal{V})$ and $\sinh(\mathcal{V})$ is recovered with $\mathcal{V}(= \tanh^{-1}(v))$.
\par 
A general Lorentz transformation can be written as an exponential containing the 
sum of a rotation and a boost, which to first order is equal to the product of a boost with a rotation. The calculations of the second and third order terms show
that the equations for the generators used in this paper, allow to reliably
infer the expressions for the higher order generators, without having recourse 
to the commutation relations.
\par
The transformation matrices for Weyl spinors are derived for finite values of the rotation and velocity, and field representations, leading to the expression for the angular momentum operator, are studied, as well as the Lorentz transformation properties of the generators.

\smallskip
\smallskip \textbf{1. Rotations and Boosts. Arbitrary Axes.}
\smallskip
By definition, a Lorentz transformations of coordinates, namely,
\begin{equation}\label{E:int1}
x'^\mu = \Lambda^\mu_{\mspace{8mu}\nu}x^\nu =\left(\delta^\mu_\nu+\omega^\mu_
{\mspace{8mu}\nu}\right)x^\nu
\end{equation}
leaves the metric,
\begin{equation}
ds^2 = g_{\mu\nu}dx^\mu dx^\nu = dt^2 - dx^2 - dy^2 - dz^2,
\end{equation}
invariant. It is straightforward to show that this condition imposes the following
constraints on the coefficients $\Lambda^\mu_{\mspace{8mu}\nu},$
\begin{equation}
g_{\mu\nu}=g_{\alpha\beta}\Lambda^\alpha_{\mspace{8mu}\mu}\Lambda^\beta_
{\mspace{8mu}\nu}, {\mspace{50mu}}\omega_{\mu\nu} + \omega_{\nu\mu}=0
\end{equation}
where the second equation is the linear version of the first one for small
values of $\omega_{\mu\nu} =g_{\mu\alpha}(\Lambda^\alpha_{\mspace{8mu}\nu}
-\delta^\alpha_{\mspace{8mu}\nu})$.
The axis of a rotation by an angle $\Theta,$ counter clockwise is positive,
will be denoted by $\Vr = (\rx,\ry,\rz)$ = $(\rone,\rtwo,\rthree)$,
a unit vector. The matrix associated with this Lorentz transformation is
($\Omega = 1 - \cos \Theta$),
\begin{equation}
\VLambda_{rot}(\Theta,\Vr) =
\left(
\begin{matrix}
1&0&0&0                          \\
0&\rx^2+(1-\rx^2)\cos\Theta             &
   \rx\ry\Omega-\rz\sin\Theta           & 
   \rx\rz\Omega+\ry\sin\Theta     \\
0&\rx\ry\Omega+\rz\sin\Theta            &
  \ry^2+(1-\ry^2)\cos\Theta             &   
 \ry\rz\Omega\!-\!\rx\sin\Theta    \\ 
0&\rx\rz\Omega-\ry\sin\Theta            &
\ry\rz\Omega+\rx\sin\Theta              &   
 \rz^2+(1\!-\rz^2)\cos\Theta         
\end{matrix}
\right )
\end{equation}
The matrix associated with a boost with a velocity $v$ and a direction defined by the unit vector $\Vb = (\bx,\by,\bz)$ = $(\bone,\btwo,\bthree)$
is given by, ( $\gamma = (1 - v^2)^{-1/2} $, $ c$ has been set equal
to one ), 
\begin{equation}
\VLambda_{boost}(v,\Vb)=
\left(
\begin{matrix}
\gamma     &  v\bx\gamma         &  v\by\gamma        & v\bz\gamma\\
v\bx\gamma & 1+(\gamma-1)\bx^2 &  (\gamma-1)\bx\by & (\gamma-1)\bx\bz\\
v\by\gamma & (\gamma-1)\by\bx  & 1+(\gamma-1)\by^2 & (\gamma-1)\by\bz\\
v\bz\gamma & (\gamma-1)\bz\bx  & (\gamma-1)\bz\by & 1+(\gamma-1)\bz^2
\end{matrix}
\right)
\end{equation}

\smallskip
\smallskip \textbf{2. Generators and Commutation Relations}.
\smallskip
It follows from Eqs.(4) and (5), that the matrix for a Lorentz transformation, combining a boost and a rotation, and linear in $v$ and $\Theta$, is given by,
\begin{equation}
\VLambda = 1 +
\left(
\begin{matrix}
0    &     v\bone &     v\btwo & v\bthree     \\
v \bone &        0 & -\gth\rthree & \gth\rtwo  \\
v\btwo &  \gth\rthree &        0 & -\gth\rone \\
v\bthree & -\gth\rtwo &  \gth\rone & 0        \\
\end{matrix}
\right ) 
= 1+\Theta G_{rot}(\Vr)\ + v G_{boost}(\Vb)
\end{equation}
where,
\begin{equation}
 G_{rot}(\Vr) =
\left(
\begin{matrix}
0    &     0 &     0 & 0     \\
 0 &        0 & -\rthree & \rtwo  \\
0 &  \rthree &        0 & -\rone \\
0 & -\rtwo &  \rone & 0        \\
\end{matrix}
\right ),
\end{equation} 
and
\begin{equation}
 G_{boost}(\Vb)= 
\left(
\begin{matrix}
0    &     \bone &     \btwo & \bthree     \\
 \bone &        0 & 0 & 0  \\
\btwo &  0 &   0 & 0 \\
\bthree & 0 &  0& 0        \\
\end{matrix}
\right ) 
\end{equation}
are the generators for a rotation around $\Vr$, and for a boost along $\Vb$, respectively. In the right-hand side of Eq.(6), $\Theta$ and $v$ should be interpreted as diagonal matrices. From Eqs.(7) and (8) one finds that  the 
commutation rules for the rotation-rotation generators are,
\begin{equation}
\lbrack \Gr,\Grp \rbrack =
\left(
\begin{matrix}
0 & 0 &  0 & 0 \\
0 & 0 &  \rtwo\rxp -\rone\ryp & \rthree\rxp - \rone\rzp \\ 
0 & \rone\ryp -\rtwo\rxp &  0 & \rthree\ryp -\rtwo\rzp \\
0 & \rone\rzp - \rthree\rxp  & \rtwo\rzp -\rthree\ryp) & 0 
\end{matrix}
\right)
\end{equation}
which obviously defines a rotation. The coefficients of this matrix are
the components of the vector $\Vr\wedge\Vrp$ and clearly, 
\begin{equation}
\lbrack \Gr,\Grp \rbrack  = G_{rot}(\Vr\wedge\Vrp) 
\end{equation}
For the boost-boost generators one obtains,
\begin{equation}
\lbrack \Gb,\Gbp \rbrack =
\left(
\begin{matrix}
0 & 0 &  0 & 0 \\
0 & 0 &  \bone\byp -\btwo\bxp & \bone\bzp - \bthree\bxp \\ 
0 & \btwo\bxp -\bone\byp &  0 & \btwo\bzp -\bthree\byp \\
0 & \bthree\bxp - \bone\bzp  & \bthree\byp -\btwo\bzp) & 0 
\end{matrix}
\right)
\end{equation}
also a rotation, and
\begin{equation}
\lbrack \Gb,\Gbp \rbrack  = -G_{rot}(\Vb\wedge\Vbp)
\end{equation}
For the rotation-boost, the commutation relations are, 
\begin{equation}
\lbrack \Gr,\Gb \rbrack =
\left(
\begin{matrix}
0&\rtwo\bthree-\rthree\btwo &\rthree\bone-\rone\bthree &\rone\btwo-\rtwo\bone\\
\rtwo\bthree-\rthree\btwo   & 0 &  0 & 0 \\ 
\rthree\bone-\rone\bthree & 0 &  0 & 0 \\
\rone\btwo-\rtwo\bone & 0 &  0 & 0 
\end{matrix}
\right)
\end{equation}
\begin{equation}
\lbrack \Gr,\Gb \rbrack   =  G_{boost}(\Vr\wedge\Vb)
\end{equation}
As expected the commutation relations in Eqs. (10), (12), and (14) are in agreement with those satisfied by the $J^i$ and $K^i$ generators derived from a Lorentz transformation as follows,
\begin{equation}\nonumber
x'^\alpha=\left(\delta^\alpha_\beta+\omega^\alpha_{\mspace{8mu}\beta}\right)
x^\beta =\left(\delta^\alpha _\beta+g^{\alpha\mu}\omega_{\mu\beta}\right) x^\beta =
\left(\delta^\alpha_\beta+g^{\alpha\mu}\delta^\nu_\beta{\mspace{2mu}} \omega_{\mu\nu}\right) x^\beta 
\end{equation}
\begin{equation}
= \left(
\delta ^\alpha _\beta +\frac{1}{2}\omega_{\mu\nu}
\left(g^{\alpha\mu}\delta^\nu_\beta- g^{\alpha\nu}\delta^\mu_\beta\right)
\right) x^\beta  
=\left(\delta ^\alpha _\beta +
\frac{1}{2}\omega_{\mu\nu}\left(J^{\mu\nu}\right)^\alpha_{\mspace{8mu}\beta}\right)
x^\beta 
\end{equation}
In Eq.(15), the fifth term appeals to the linear Lorentz condition, $\omega_{\mu\nu}+\omega_{\nu\mu} = 0$, followed by a relabeling ($\mu \rightleftarrows \nu$), and the sixth term shows that the expression for 
the generators is given by,
\begin{equation}
\left(J^{\mu\nu}\right)^\alpha_{\mspace{8mu}\beta} = 
g^{\alpha\mu}\delta^\nu_\beta - g^{\alpha\nu}\delta^\mu_\beta
\end{equation}
Define,
\begin{equation}
J^i = \frac{1}{2}\eps J^{jk},\mspace{40mu}K^i = J^{io},\mspace{40mu}
\theta^i = \frac{1}{2}\eps\omega^{jk},\mspace{40mu}\eta^i=\omega^{io}
\end{equation}
Then, $J^i$, $K^i$ satisfy the same commutation rules (namely Eqs.(10), (12)
and (14)), than $\Gr$ and $\Gb$ with $\Vr$ and $\Vb$ being equal to
$\Veone $, $\Vetwo $, $\Vethree$, the unit basis vectors. It is now straightforward
to show that,
\begin{equation}
x'^\alpha =\left(\delta ^\alpha _\beta +
\frac{1}{2}\omega_{\mu\nu}\left(J^{\mu\nu}\right)^\alpha_{\mspace{8mu}\beta}\right)
x^\beta = \left(\delta ^\alpha _\beta - \left (\vec J \cdot \boldsymbol {\theta} - 
\vec K \cdot \boldsymbol{\eta} \right)^\alpha_{\mspace{8mu}\beta}\right)x^\beta 
\end{equation}
In this equation the first minus sign would be a plus one, if the metric coefficients, defined in Eq.(2), are reversed.
The matrix for the Lorentz transformation defined in Eq.(18) is 
found to be,
\begin{equation}
 \Lambda = 1 +
\left(
\begin{matrix}
0    &  \etone  &  \ettwo  & \etthree   \\
\etone  &   0   & -\tthree  &  \ttwo  \\
\ettwo  &  \tthree  &    0  & -\tone  \\
\etthree  & -\ttwo  &  \tone  &   0   \\
\end{matrix}
\right )
\end{equation} 
A comparison with Eq.(6) reveals that $v\bi =\eti$ and $\Theta\ri= \ti $
Because $\Vb$ and $\Vr$ are unit vectors it follows that,
\begin{equation}
v = \pm \left( (\etone)^2 + (\ettwo)^2 + (\etthree)^2\right )^{1/2},\mspace{15mu}
\Theta = \pm\left((\tone)^2 + (\ttwo)^2 + (\tthree)^2\right )^{1/2}
\end{equation}
and in consequence,
\begin{equation}
\ri =\pm \ti/{\mid \boldsymbol\theta \mid} \mspace{40mu}
\bi =\pm \eti/{\mid\boldsymbol\eta \mid}
\end{equation}
Therefore $\ti, \eti$ {\em determine the axis of rotation and
boost, but not the sense of rotation nor the direction of the boost}

\smallskip
\smallskip \textbf{3. Finite rotations and boosts}.
\smallskip
The matrix for a general Lorentz transformation is usually written as,
\begin{equation}
\VLambda = \exp\{-\vec J \cdot \boldsymbol {\theta} + 
\vec K \cdot \boldsymbol{\eta} \}
\end{equation}
Working with the generators defined in Eqs.(7) and (8),
this expression for $\VLambda$ is replaced by,
\begin{equation}
\VLambda = \exp\{\Theta G_{rot}(\Vr)\ + v G_{boost}(\Vb)\},
\end{equation}
Eq.(23) is reminiscent of a Taylor expansion for a function of two
variables, namely,
\begin{equation}
f(\Theta,v) = \exp \{\Theta \pTheta + v\pv \}f(\Theta,v)
\end{equation}
where the derivatives play the role of the generators. The derivatives, unlike
generators, do however commute.
\smallskip

a) {\sl Rotations}. It needs to be shown that $\VLambda_{rot}(\Theta,\Vr)$,
as given by Eq.(4), is equal to, 
\begin{equation}
\exp\{\Theta G_{rot}(\Vr)\} = \mspace{3mu}\VLambda = 1 + \Theta\Gr + \frac{1}{2} \Theta^2 \Gr^2 + \frac{1}{3!}\Theta^3\Gr^3 + ......
\end{equation} 
The relabeling $\VLambda_{rot}(\Theta,\Vr)= \VLambda$ is for notational convenience.
For the above expansion to be valid it is necessary that,
\begin{equation}
\pTheta \VLambda = \Gr\VLambda
\end{equation}
From Eq.(4) it is found that,
\begin{equation}
\pTheta \VLambda =
\left(
\begin{matrix}
0&0&0&0                          \\
0&-(1-\rx^2)\sin\Theta             &
   \rx\ry\sin\Theta-\rz\cos\Theta           & 
   \rx\rz\sin \Theta+\ry\cos\Theta     \\
0&\rx\ry\sin \Theta+\rz\cos\Theta            &
  -(1-\ry^2)\sin\Theta                   &   
 \ry\rz\sin \Theta\!-\!\rx\cos\Theta    \\ 
0&\rx\rz\sin \Theta-\ry\cos\Theta            &
\ry\rz\sin \Theta+\rx\cos\Theta              &   
 -(1-\rz^2)\sin\Theta         
\end{matrix}
\right )
\end{equation} 
and it is straightforward to verify that Eq.(26) is indeed satisfied. 
Therefore Eq.(25) is the correct expression for finite rotations.
\par
The expression for $\VLambdar(\Theta,\Vr)$ in Eq.(4) can be derived from {\em just the first three terms} of the expansion for $\VLambdar(\Theta,\Vr)$ in 
Eq.(25): the term in $\Theta$ is replaced by $\sin\Theta$ and the next one, in $\frac{1}{2}\Theta^2$, by $\Omega$ (notice that
$\Omega = 1-\cos\Theta = \frac{1}{2}\Theta^2 + \mathcal{O}(\Theta^4)$). How is this 
possible?, an alert reader is bound to ask, $\Theta$ does not stand in isolation, instead it is multiplied by $\Gr$, which surely cannot be ignored. Very good
 question indeed, that allows  however for an apparent answer:

\newpage
\begin{equation}\nonumber
\Gr = -\Gr^3 = \Gr^5 = -\Gr^7 =....,{\mspace{15mu}}  \Gb = \Gb^3 = 
\end{equation}
\begin{equation}\nonumber
\Gb^5 = \Gb^7 =..,{\mspace{15mu}} \Gr^2 = -\Gr^4 = \Gr^6 = -\Gr^8 = ..
\end{equation}
\begin{equation}
\Gb^2 = \Gb^4 = \Gb^6 = \Gb^8 =.....
\end{equation}
It is important to keep in mind that Eq.(26) implies the differential version of the addition of rotations, namely, 
$\VLambdar(\Theta + d\Theta)=\VLambdar(\Theta)
\VLambdar(d\Theta)$. Indeed,
\begin{equation}\nonumber
\VLambdar(\Theta + d\Theta) = \VLambdar(\Theta) + d\Theta \pTheta \VLambdar(\Theta)=
\VLambdar(\Theta) + d\Theta \Gr\VLambdar(\Theta) =
\end{equation} 
\begin{equation}
=(1+ d\Theta \Gr)\VLambdar(\Theta) = \VLambdar(d\Theta)\VLambdar(\Theta)
\end{equation}

\smallskip

b) {\sl Boosts}.
No expansion for $\VLambdab$ analogous to the one for $\VLambdar$, i.e. with
 $\Theta \rightarrow v$ and $\Gr \rightarrow \Gb$, can exist, because as shown above, it would imply that
$\VLambdab(v + dv)=\VLambdab(b)\VLambdab(db)$, in contradiction with the non additive property of velocities in Relativity. Assume therefore instead that
an expansion of the form (in the text of this paper, and contrary to the Abstract,
the symbol $\mcV$ will designate a function of the velocity, $v$).
\begin{equation}
\mcV = \mcV(v),\mspace{10mu} \VLambdab(\mcV) = \exp\{\mcV G_{boost}(\Vb)\} = 1 + \mcV\Gb + \frac{1}{2} \mcV^2 \Gb^2 +  .....
\end{equation}
Then,
\begin{equation}
\pmcV \VLambdab(\mcV) = \VLambdab (\mcV) \Gb
\end{equation}
is satisfied. Imposing this condition on $\VLambdab=\VLambdab(\Vb,v)$, 
as given by Eq.(5), leads to the following relations,
\begin{equation}
\frac{d\gamma}{d\mcV} = {v}\gamma, {\mspace{50mu}}  \frac{d(v \gamma)}{d\mcV} = \gamma
 = (1 - v^2)^{-1/2},
\end{equation}
It is now straightforward to derive an equation for $\pv \mcV({\rm v})$, 
\begin{equation}\nonumber
\frac {d\gamma}{d\mcV}= \frac {d(1 - v^2)^{-1/2}}{d\mcV} =
\gamma^2\gamma v{\mspace{4mu}}\frac{dv}{d\mcV}=
 \gamma^2\frac {d\gamma}{d\mcV}{\mspace{4mu}}\frac{dv}{d\mcV}\rightarrow 
1 =\gamma^2 \frac{dv}{d\mcV} \longrightarrow 
\end{equation}
\begin{equation}
\frac{d\mcV}{dv}= \frac{1}{1-v^2} = \frac { d }{dv} {\rm{tanh}^{-1}} v  
\end{equation}
Therefore,
\begin{equation}
\gamma = \cosh(\mcV),{\mspace{20mu}}
\mcV(v) = \tanh^{-1}(v) = v + v^3/3 + v^5/5 +......
\end{equation}
The matrix for the boost, namely $\VLambdab(v,\vec b)$ in Eq.(5), can now be written,
\begin{equation}
\VLambda_{boost}(\mcV,\Vb)=
\left(
\begin{matrix}
\gamma          & \bx \sinh(\mcV)   &  \by\sinh(\mcV)   &  \bz\sinh(\mcV)\\
\bx\sinh(\mcV)  & 1+(\gamma-1)\bx^2 & (\gamma-1)\bx\by  & (\gamma-1)\bx\bz\\
\by\sinh(\mcV)  & (\gamma-1)\by\bx  & 1+(\gamma-1)\by^2 & (\gamma-1)\by\bz\\
\bz\sinh(\mcV)  & (\gamma-1)\bz\bx  & (\gamma-1)\bz\by  & 1+(\gamma-1)\bz^2
\end{matrix}
\right)
\end{equation}
It is again possible to derive the expression for $\VLambdab(v,\Vb)$,
(cf. Eq.(5), and Eq.(28)) from the first three terms of Eq.(30). 
In the linear term of this expansion, replace $\mcV$ by 
$v\gamma$   (notice that $v\gamma= d\gamma/d \mcV =d\cosh (\mcV)/d\mcV = 
\sinh \mcV =  \mcV + \mathcal{O}(mcV^3)$), and in the quadratic term, replace $\mcV^2/2$, by $\gamma -1$ ( $\gamma-1 = \mcV^2/2 + \mathcal{O}(\mcV^4) $. The linear replacement reproduces the first line and the first column of   
 Eq(5), whereas the quadratic replacement introduces the factor $\gamma -1$ present in $\VLambdab(v,\Vb)$. 
\smallskip

c) {\sl Rotations and Boosts}.
Given the Lorentz transformation, $\boldsymbol{\mathcal{L}} \boldsymbol{\mathcal{T}}$ 
 = \\
$\exp\{\Theta G_{rot}(\Vr) + \mcV G_{boost}(\Vb)\}$, the problem at hand 
 is to find the product of rotations and boosts equivalent to $\boldsymbol{\mathcal{L}} \boldsymbol{\mathcal{T}}$. To first order in $\Theta$ and $\mcV$, the transformation
$\boldsymbol{\mathcal{L}} \boldsymbol{\mathcal{T}}$ and 
 $ \exp\{\Theta G_{rot}(\Vr)\}\times \exp \{\mcV G_{boost}(\Vb)\}$ are of course equal, but to second order, the terms in $\Theta \mcV$ differ. Therefore a relation of the following form must exist,
\begin{equation}
\exp\{\Theta \Gr + \mcV \Gb\} =
\exp\{\Theta \Gr\}\times \exp \{\mcV\Gb \}\times
\exp\{\alpha \Theta \mcV {\mspace{3mu}}G_x(\Vv)\}
\end{equation}
where $\alpha$ is a constant to be determined, $x$ stands for a boost or
rotation, and $\Vv$ for the unit vector that defines the axis of 
rotation or the boost's direction. The expression for $G_x(\Vv)$ can be easily
inferred: $\Vv$ must be a combination of the vectors $\Vr$ and $\Vb$ and the obvious choice is $\Vv = \Vr\wedge\Vb$, which defines a boost, as Eq.(14)
shows. Therefore $G_x(\Vv) = \Gbb(\Vr\wedge\Vb)$. The equality of the terms in
$\Theta \mcV$ in Eq.(36) leads then to,
\begin{equation}
\frac{1}{2}\left( G_{rot}G_{boost} + G_{boost}G_{rot}\right) =
G_{rot}G_{boost} + \alpha \mspace{2mu}G_{boost}(\Vr\wedge\Vb)
\end{equation}
which can indeed be satisfied with $\alpha = -1/2$. The notation (to be used
from now on) has been simplified: $G_{rot} = \Gr$ and $ G_{boost} = \Gb$.
Eq.(37) can also be written, 
\begin{equation}
-\frac {1}{2}[\Grb, \Gbb] = \alpha G_x(\Vv)
\end{equation}
If no prior knowledge of $G_x(\Vv)$ had been assumed, this equation would have 
determined $\alpha$ and $G_x(\Vv)$, which, because these are the commutation relations, {\em cannot be other than} $G_{boost}(\Vr\wedge\Vb)$.
\par
It is of interest to calculate the third order terms. There are now two terms,
namely those in $\Theta^2\mcV$ and $\mcV^2\Theta$, that require to be 
balanced by two new Lorentz transformations which are determined by:
\begin{equation}\nonumber
\exp\{\Theta \Grb + \mcV \Gbb\} =
\exp\{\Theta \Grb\}\times \exp \{\mcV\Gbb \}\times
\exp \{-\frac {1}{2}\Theta \mcV \mspace{2mu}G_{boost}(\Vr\wedge\Vb)\} \times
\end{equation}
\begin{equation}
\times\exp \{ \alpha {\mspace{3mu}}\Theta^2\mcV { \mspace{3mu} } G_x(\Vv) \} \times
\exp\{\beta { \mspace{3mu} }\Theta \mcV^2{\mspace{3mu}}G_y(\Vv')\}
\end{equation}
where, $x, y$ stand for boost or rotation and $\Vv,\Vv'$ for the 
vectors defining the axes of the transformations. It is again possible
to infer the expressions for $G_x(\Vv)$ and $G_y(\Vv') $. The obvious
choices for $\Vv$ and $\Vv'$ are now $\Vv =\Vr\wedge(\Vr\wedge\Vb)$ and
$\Vv' =\Vb\wedge(\Vr\wedge\Vb)$. It is clear that $\Vv$ with two 
$\Vr$ s and one $\Vb$, should be associated with $\Theta^2\mcV$ and that $\Vv'$ with two $\Vb$s and one $\Vr$, with $\mcV^2\Theta$.
Keeping in mind that $\Vr\wedge\Vb$ is a boost, Eq.(14)
suggests that $x$ is a boost, and Eq.(12), that  $y$ is a rotation. Therefore,
\begin{equation}
 G_x(\Vv) = G_{boost}\left( \Vr\wedge(\Vr\wedge\Vb)\right){ \mspace{40mu} }
 G_y(\Vv')  = G_{rot}\left( \Vb\wedge(\Vr\wedge\Vb) \right)
\end{equation}
The equality of the terms in $\Theta^2\mcV$ in Eq.(39) leads then to,
\begin{equation}\nonumber
\frac {1}{6}\left(\Grb^2 \Gbb + \Grb\Gbb\Grb +\Gbb\Grb^2\right)= 
\frac{1}{2}\left( \Grb^2\Gbb - \Grb G_{boost}(\Vr\wedge\Vb)\right)
\end{equation}

\begin{equation}
 +\alpha G_{boost}\left( \Vr\wedge(\Vr\wedge\Vb)\right)
\end{equation}
If $\Vr =\vec e_z$ and $\Vb = \vec e_x$, then  $\Vr\wedge\Vb $ =
$\vec e_y$ and $\Vr\wedge (\Vr\wedge\Vb)$ = $-\vec e_x$. All the {\em three}
dimensional matrices in Eq.(41) are then easily evaluated, and 
the value found for $\alpha$ is $1/6$. Similar calculations for the terms in
$\Theta\mcV^2$  show that $\beta = 1/3$. If no prior knowledge for $G_x(\Vv)$
is assumed, $G_x(\Vv)$ is of course {\em determined} by the commutation relations from Eq.(41). Contrast this statement with the one appropriate to the second order terms, where $G_x(\Vv)$ is a {\em member} of the commutation relations.
\par
It is clear that the Lorentz transformation $ \exp\{\Theta G_{rot}(\Vr) + \mcV G_{boost}(\Vb)\}$, is a complex product of rotations and boosts, not shedding light
(as $ \exp\{\Theta G_{rot}(\Vr)$ and $\mcV G_{boost}(\Vb)\}$ do), on finite transformations. 

There can be little doubt that the theory of group representations could be developped as well from  the product of a boost and a rotation, namely
$ \exp\{\Theta G_{rot}(\Vr)\}\times \exp\{\mcV G_{boost}(\Vb)\}$.

\smallskip
\smallskip \textbf{4. Spin}.
\smallskip
Define,
\begin{equation}
J^{\pm}(\vec u) = i (\Grb(\vec u) \pm i\Gbb(\vec u))/2 = (i\Grb(\vec u) 
\mp \Gbb(\vec u))/2
\end{equation}
The commutation rules are given by,
\begin{equation}\nonumber
[J^{+}(\vec u),J^{+}(\vec u')] = i J^{+}(\vec u \wedge \vec u')
\end{equation}
\begin{equation}\nonumber
[J^{-}(\vec u),J^{-}(\vec u')] = i J^{-}(\vec u \wedge \vec u')
\end{equation}
\begin{equation}
[J^{+}(\vec u),J^{-}(\vec u')] = 0
\end{equation}
With the help of Eqs. (7) and (8), the calculation  of $(J^{\pm}(\vec u))^2$
is straightforward. It is found that ($\bf{I}$ is the unit matrix),
\begin{equation}
(J^{\pm}(\vec u))^2 = {\bf {I}}/4,
\end{equation}
The two degrees of freedom associated with particles of spin 1/2 suggests 
working with a representation for the generators on a vector space of dimension two. The Pauli matrices are given by,
\begin{equation}\nonumber
\sigma^1 =\left(
\begin{matrix}
0&1\\
1&0
\end{matrix}
\right)\mspace{25mu}
\sigma^2=\left(
\begin{matrix}
0&-i\\
i&0
\end{matrix}
\right)\mspace{25mu}
\sigma^3 =\left(
\begin{matrix}
1&0\\
0&-1
\end{matrix}
\right)
\end{equation}
\begin{equation}
M(\vec u) = \Vsigma \cdot \vec u = \left(
\begin{matrix}
u_z&u_x -i u_y\\
u_x + i u_y&-u_z
\end{matrix}
\right)\mspace{50mu} M^2(\vec u) = {\bf {I}}
\end{equation}
The commutation relations for the matrix $M(\vec u)$, defined above, are given by,
\begin{equation}
[M(\vec u),M(\vec u')]= 2i
\left(
\begin{matrix}
U_z & U_x -iU_y \\
U_x +iU_y & -U_z
\end{matrix}
\right)
\mspace{25mu}
\vec U = \vec u \wedge \vec u'
\end{equation}
Define
\begin{equation}
\GsruL = \frac{1}{2i}M(\vec u)\mspace{30mu} \GsbuL = \frac{1}{2}M(\vec u)
\mspace{30mu}{\cal J}^{+}(\vec u) = 0
\end{equation}
\begin{equation}
\GsruR = \frac{1}{2i}M(\vec u)\mspace{30mu} \GsbuR = -\frac{1}{2}M(\vec u)
\mspace{30mu}{\cal J}^{-}(\vec u) = 0
\end{equation}
where the ${\cal J}^{\pm}$ are defined as in Eq.(42) with the ${\cal G}$ generators
replacing the $G$ ones. With the help of Eq.(46) it is straightforward to show that 
$\GsruL,\mspace{1mu} \GsbuL$ (and of course $\GsruR,\mspace{1mu} \GsbuR$ as well), obey the same commutation relations (namely Eqs. (10), (12), and  (14)), than $\Grb(\vec u)$ and $\Gbb(\vec u)$. Eqs. (47) and (48) are two representations of
the Lorentz group, known as the left- and right handed Weyl spinors, that can be labelled by $(({\cal J}^{-})^2 = \frac{1}{2},\mspace{2mu} {\cal J}^{+} = 0 )$ and
 $({\cal J}^{-} = 0, ({\cal J}^{+})^2 = \frac{1}{2} )$ respectively.
It is of interest to calculate the series
\begin{equation}\nonumber
\VLambda^L(\Theta,\vec u) = 1 + \Theta\GsruL  + 
\frac{1}{2}\Theta^2\GsruL^2  + ..............
\end{equation}
\begin{equation}
\VLambda^L(\mcV,\vec u) = 1 + \mcV \GsbuL^2 + 
\frac{1}{2}\mcV^2\GsbuL^2  + ..............
\end{equation}
Because the square of the matrix in the expressions for $\GsruL$ and 
$\GsbuL $ is equal to the unit matrix, the calculations are straightforward. it
it is found,
\begin{equation}
\VLambda^L(\Theta,\vec u) =
\left(
\begin{matrix}
\cos(\Theta/2) - iu_z\sin(\Theta/2) &-i\sin(\Theta/2)(u_x -iu_y)\\
-i\sin(\Theta/2)(u_x + i u_y) & \cos(\Theta/2) +i u_z\sin(\Theta/2)
\end{matrix}
\right)
\end{equation}
\begin{equation}
\VLambda^L(\mcV,\vec u)=
\left(
\begin{matrix}
\cosh(\mcV/2) +u_z \sinh(\mcV/2) & \sinh(\mcV/2)(u_z -iu_y)\\
\sinh(\mcV/2) (u_z +iu_y)& \cosh(\mcV/2) - u_z \sinh(\mcV/2)
\end{matrix}
\right)
\end{equation}
\par
For finite rotations and boosts, these matrices define the transformations for the left handed Weyl spinor components, as $\VLambdar(\Theta,\vec r)$ and 
$\VLambdab(v,\vec b)$  (in Eqs. (4) and (35)), do for the coordinates of a Lorentz transformation. For the right handed Weyl spinor, $\VLambda^R(\Theta,\vec u)$ remains unchanged, whereas $\mcV$ is replaced by $-\mcV$ in the expression for $\VLambda^R(\mcV,\vec u)$.

\smallskip
\smallskip \textbf{4. Field Representations}.
\smallskip
Let $T^{\mu\nu}$ ($\mu$ is line, and $\nu$ is column) be the antisymmetric tensor associated with the vectors $\vec r$ and $\vec b$ $( T^{0i}= b^i, T^{i0} -b^i$ and the rotational
components as in Eq.(7)), then
\begin{equation}
T{^\mu\mspace{1mu}_{\nu}} = -\left(
\begin{matrix}
0   & b^1  &  b^2       & b^3      \\
b^1 & 0    & -\rthree   & \rtwo  \\
b^2 &  \rthree &      0 & -\rone \\
b^3 & -\rtwo &  \rone & 0        \\
\end{matrix}
\right)
\end{equation}
Indeed, because of Eq.(2), $T^{0}{\mspace{1mu}_{i}}= g_{i \alpha}T^{0 \alpha} = g_{i i} T^{0 i} = - T^{0 i}= - b^{i}$. On the other hand, $T^{i}{\mspace{1mu}}_{0} =
g_{0\alpha}T^{i\alpha}=T^{i 0} = -b^i $, i.e., the boost components of $T^{\mu \nu}$ become symmetric, whereas the rotation components remain antisymmetric but change sign. The matrix in Eq.(52) is equal to the sum of $\Gr$ and $\Gb$ as given by Eqs. (7) and (8).\\
We define therefore the following operators,
\begin{equation}
A_{rot}(\vec r, x) = -{\Gr}^{\mu}_{\mspace{7mu}\nu}
{\mspace{2mu}} x^{\nu }\partial_{\mu} \mspace{50mu}
A_{boost}(\vec b, x) = -{\Gb}^{\mu}_{\mspace{7mu}\nu}
{\mspace{2mu}} x^{\nu }\partial_{\mu}
\end{equation}
The proof that $A_{rot}(\vec r, x)$ and $A_{boost}(\vec b, x)$ satisfy the same
commutation rules than $\Gr$ and $\Gb$, namely Eqs.(10), (12), (14), is straightforward; it appeals to the relations, $ \Gr ^{\mu}_{\mspace{7mu}\nu} =
 -\Gr ^{\nu}_{\mspace{7mu}\mu}$, and $ \Gb ^{\mu}_{\mspace{7mu}\nu} =  \Gb^{\nu}_{\mspace{7mu}\mu}$, and involves some relabeling. It should be stressed
that the minus signs in Eqs. (53), essential for this proof, can be firmly justified. The explicit expressions for 
$A_{rot}(\vec r, x)$ and $A_{boost}(\vec b, x) $ are $(t, x, y, z = x^{\mu}$),
\begin{equation}\nonumber
A_{rot}(\vec r, x)= - \big( r_{x}(y\partial_{z} - z\partial_{y})
 + r_{y}(z\partial_{x} - x\partial_{z})
 + r_{z}(x\partial_{y} - y\partial_{x})\big)
\end{equation}
\begin{equation}
A_{boost}(\vec b, x) = - \big( b_{x}(x\partial_{t} + t\partial_{x})
 + b_{y}(y\partial_{t} + t\partial_{y})
 + b_{z}(z\partial_{t} + t\partial_{z})\big)
\end{equation}
The quantities that multiply the $r_i ,b_j$ in Eq.(54) define the tensor $L^{\mu}_{\mspace{7mu}\nu}$, namely,
\begin{equation}
L^{j}_{\mspace{7mu}k} = -(x^j \partial_{k} - x^k \partial_{j})\mspace{50mu}
L^{0}_{\mspace{7mu}i} = -(t \partial_i + x^i \partial_{t})
\end{equation}
Raising the lower index we obtain,
\begin{equation}
L^{\mu\nu} = x^{\mu} \partial^{\nu} - x^{\nu} \partial^{\mu}
\end{equation}
(c.f. Maggiore, 2005, p.30, Eq.2.78). From Eq.(54) it is clear that, say
$x\partial^{y} - y\partial^{x}$, is the $z$-component of the angular momentum.
The angular momentum operator acts on a function $\psi(x)$ as follows 
$(\delta x^\mu = \Theta{\Gr}^{\mu}_{\mspace{7mu}\nu}{\mspace{2mu}} x^{\nu }$),
\begin{equation}
(1 +\Theta A_{rot}(\vec r, x)) \psi(x) = 
\psi( x^\mu) -\delta x^\mu\partial_\mu\psi(x^\mu)
=\psi( x^\mu - \delta x^{\mu }) = \psi'(x^\mu)
\end{equation}
i.e. $\psi(x^\mu) \longrightarrow \psi'(x^\mu)$ and clearly $\psi(x^\mu) =
\psi'(x'^\mu)$ where $x'^\mu = x^\mu + \delta x^\mu$. The base space for the angular momentum representation is the set of scalar functions.

\smallskip
\smallskip \textbf{5. Lorentz's Transformations of the Generators}.
\smallskip
It is of interest to study the Lorentz transformations properties of the 
generators $\Grb(\vec r)$ and $\Gbb(\vec b),$ namely expressions as
$\VLambdab(\vec b) \Grb(\vec e_z) \VLambdab^{-1}(\vec b)$ and
$\VLambdar(\vec r)\Gbb(\vec e_z) \VLambdar^{-1}(\vec r),$ c.f., Weinberg, 2005, p.60
(notice that here only rotations are associated with unitary operators). It is found,
\begin{equation}\nonumber
\VLambdab(\vec b) \Grb(\vec e_z) \VLambdab^{-1}(\vec b) =  \Grb(\vec e_z)
- (\gamma-1)\Bigl( b_x b_z \Grb(\vec e_x) +  b_y b_z \Grb(\vec e_y)+
\end{equation}
\begin{equation}
(b^2_z -1)\Grb(\vec e_z)\Bigr)+v\gamma \Gbb(\vec{b}\wedge\vec{e_z})=\gamma\Grb(\vec{e_z})
 - (\gamma-1) \Grb(b_z\vec b \cdot \vec e) + v\gamma \Gbb(\vec{b}\wedge\vec{e_z})
\end{equation}
where $v$ is the boost velocity and $\gamma = (1-v^2)^{-1/2}.$
\begin{equation}\nonumber
\VLambdar(\vec r)\Gbb(\vec e_z) \VLambdar^{-1}(\vec r) =  \Gbb(\vec e_z)+ \Gbb(\vec e_x)(r_x r_z\Omega +r_y\sin\Theta) + 
\end{equation}
\begin{equation}\nonumber
+\Gbb (\vec e_y)( r_y r_z\Omega -r_x\sin \Theta) +\Gbb(\vec e_z) (r^2_z -1)\Omega =
\end{equation}
\begin{equation}
=(1 -\Omega)\Gbb(\vec e_z)+\Gbb(\Omega r_z \vec r \cdot \vec e) +\Gbb (\sin \Theta \mspace{2mu}\vec r \wedge \vec r_z)
\end{equation}
where $ \Omega = 1 - \cos \Theta.$
In the above equations, $\vec e $ is a unit vector, and the notation should be transparent: $\Gbb(\vec{b}\wedge\vec{e_z})=b_y \Gbb(\vec e_x) - b_x \Gbb(\vec{e_y)}$ for example.
Concerning Eq.(58) notice that it reduces to $\VLambdab(\vec b) \Grb(\vec e_z) 
\VLambdab^{-1}(\vec b) =  \Grb(\vec e_z)$ if $v =0,$ because $\gamma$ is then equal
to one, and also when $b_ x = b_y = 0$ because then $b_z = 1,$ i.e. the boosts are along
the axis of rotation. In Equation (59) as well, $\Gbb(\vec e_z)$ is left unchanged if
 $\Theta$ vanihes and also when the axis of the rotations is the $z-$axis.

\smallskip
\smallskip \textbf{5. Acknowledgements.}
\smallskip

I am grateful to Dr. Maggiore for valuable comments.

\smallskip
\smallskip \textbf{6. References.}
\smallskip

Maggiore, M.: {\it A Modern Introduction to Quantum Field Theory}, Oxford
University Press, New York, (2005)

\smallskip

Weinberg, S.: {\it The Quantum Theory of Fields} Vol 1, Cambridge University Press,
New York, (2005)
\end{document}